\documentclass[prb,twocolumn,floatfix]{revtex4}

\usepackage{graphicx}
\usepackage{amsmath}
\usepackage{amssymb}
\usepackage[colorlinks=true,citecolor=blue,linkcolor=blue]{hyperref}
\usepackage{amsfonts}
\usepackage{color,xcolor}
\usepackage{epstopdf}
\usepackage{braket}
\usepackage{bm}
\usepackage{bbold}

\renewcommand{\vec}[1]{\ensuremath{\boldsymbol{#1}}}

\begin{document}

\title{Spectrum of exciton states in monolayer transition metal dichalcogenides: angular momentum and Landau levels}
\date{\today}
\author{M. Van der Donck}
\email{matthias.vanderdonck@uantwerpen.be}
\affiliation{Department of Physics, University of Antwerp, Groenenborgerlaan 171, B-2020 Antwerp, Belgium}
\author{F. M. Peeters}
\email{francois.peeters@uantwerpen.be}
\affiliation{Department of Physics, University of Antwerp, Groenenborgerlaan 171, B-2020 Antwerp, Belgium}

\begin{abstract}
A four-band exciton Hamiltonian is constructed starting from the single-particle Dirac Hamiltonian for charge carriers in monolayer transition metal dichalcogenides (TMDs). The angular part of the exciton wave function can be separated from the radial part, in the case of zero center of mass momentum excitons, by exploiting the eigenstates of the total exciton angular momentum operator with which the Hamiltonian commutes. We explain why this approach fails for excitons with finite center of mass momentum or in the presence of a perpendicular magnetic field and present an approximation to resolve this issue. We calculate the (binding) energy and average interparticle distance of different excited exciton states in different TMDs and compare these with results available in the literature. Remarkably, we find that the intervalley exciton ground state in the $\mp K$ valley has angular momentum $j=\pm1$, which is due to the pseudospin of the separate particles. The exciton mass and the exciton Landau levels are calculated and we find that the degeneracy of exciton states with opposite relative angular momentum is altered by a magnetic field.
\end{abstract}

\maketitle

\section{Introduction}

Monolayer transition metal dichalcogenides (TMDs) such as MoS$_2$, MoSe$_2$, WS$_2$, WSe$_2$, etc.\cite{mak1,splendiani,mak2,zeng,cao,ross}, lack inversion symmetry, which leads to a large direct band gap in the low-energy valleys at the corners of the hexagonal first Brillouin zone. This allows for optical excitation of exciton states\cite{korn,sallen,mak3,he,chernikov}, i.e. bound systems of an electron and a hole. Monolayer TMDs are strictly two dimensional (2D) systems and as a result the excitons in these systems are very tightly bound, i.e. they have binding energies of the order of several hundreds of meV, which is two orders of magnitude larger as compared to excitons in conventional three dimensional semiconductors\cite{elliot,kulak,expT0,francoisE,francoisT}.

In contrast to the (2D) hydrogen atom in which states with the same principal quantum number but different angular momentum are degenerate, non-local screening effects in 2D TMDs lead to the breaking of this degeneracy. The higher angular momentum states can not be observed by means of one-photon transitions which are most commonly used in experiments. Two-photon transitions can however give optical access to these states, as was shown successfully for $p$-states\cite{he,spd}. Higher order angular momentum states, such as $d$-states, have so far not been experimentally measured in 2D TMDs. Even though these non-zero angular momentum states are optically inactive, they do play an important role in exciton relaxation and valley dynamics\cite{relax}.

Studies of magnetic field effects on excitons in monolayer TMDs have mostly focused on the valley Zeeman effect\cite{valley1,valley2,valley3,valley4,valley5,valley6}, which originates mainly from the valley-contrasting magnetic moments of the valence electrons around their atomic sites. On the other hand, there is little to no work done on Landau quantization of exciton states in monolayer TMDs.

In the present paper we present a model allowing to calculate the (binding) energy, wave function, and average interparticle distance of different angular momentum exciton states in monolayer TMDs. We also calculate the exciton Landau levels and show how the magnetic field affects the degeneracy of the different states.

Our paper is organized as follows. In Sec. \ref{sec:model} we present the outline of the four-band model in which the exciton Hamiltonian and total angular momentum operator are constructed. The eigenvalue equation of this Hamiltonian is readily solved numerically for excitons with zero center of mass momentum in Sec. \ref{sec:K0} and a comparison with available experimental and theoretical results is made. For excitons with non-zero center of mass momentum an approximation is needed in order to solve this equation, as is shown in Sec. \ref{sec:Kn0}. In Sec. \ref{sec:excitonll} we calculate the exciton Landau levels and in Sec. \ref{sec:Summary and conclusion} we summarize the main conclusions.

\section{Exciton Hamiltonian and total angular momentum}
\label{sec:model}

We start from the effective low-energy single-electron Hamiltonian\cite{theory1} in the basis $\mathcal{B}^e=\{\ket{\phi^e_c},\ket{\phi^e_v}\}$ spanning the 2D Hilbert space $\mathcal{H}^e$, with $\ket{\phi^e_c}$ and $\ket{\phi^e_v}$ the atomic orbital states at the conduction $(c)$ and valence $(v)$ band edge, respectively:
\begin{equation}
\label{singelhame}
H^e_{s,\tau}(\vec{k}) = at(\tau k_x\sigma_x+k_y\sigma_y)+\frac{\Delta}{2}\sigma_z+\lambda s\tau\frac{I_2-\sigma_z}{2},
\end{equation}
where $\sigma_i$ ($i=x,y,z$) are Pauli matrices, $I_2$ is the two by two identity matrix, $a$ the lattice constant, $t$ the hopping parameter, $\tau=\pm1$ the valley index, $s=\pm1$ the spin index, $\Delta$ the band gap, and $\lambda$ the spin-orbit coupling strength leading to a spin splitting of $2\lambda$ at the valence band edge. Since a hole with wave vector $\vec{k}$, spin $s$, and valley index $\tau$ can be described as the absence of an electron with opposite wave vector, spin, and valley index, the single-hole Hamiltonian can immediately be obtained from the single-electron Hamiltonian and is given by $H^h_{s,\tau}(\vec{k})=-H^e_{-s,-\tau}(-\vec{k})$. The eigenstates of this Hamiltonian span the 2D Hilbert space $\mathcal{H}^h_{s,\tau}$. The total exciton two-body Hamiltonian acts on the product Hilbert space spanned by the tensor products of the single-particle states at the band edges\cite{twobody}, $\mathcal{B}_{\alpha}=\mathcal{B}^e_{s^e,\tau^e}\otimes\mathcal{B}^h_{s^h,\tau^h}$, and is given by
\begin{equation}
\label{excham}
\begin{split}
H^{exc}_{\alpha}(\vec{k}^e,\vec{k}^h,r_{eh}) = &H^e_{s^e,\tau^e}(\vec{k}^e)\otimes I_2 \\
&-I_2\otimes H^e_{-s^h,-\tau^h}(-\vec{k}^h)-V(r_{eh})I_{4},
\end{split}
\end{equation}
where $\alpha$ is a shorthand notation for $s^e,\tau^e,s^h,\tau^h$ and where the electron-hole interaction potential is, due to non-local screening effects, given by \cite{screening1,screening2,screening3}
\begin{equation}
\label{inter}
V(r_{ij}) = \frac{e^2}{4\pi\kappa\varepsilon_0}\frac{\pi}{2r_0}\left[H_0\left(\frac{r_{ij}}{r_0}\right)-Y_0\left(\frac{r_{ij}}{r_0}\right)\right],
\end{equation}
with $r_{ij}=|\vec{r}_i-\vec{r}_j|$, where $Y_0$ and $H_0$ are the Bessel function of the second kind and the Struve function, respectively, with $\kappa=(\varepsilon_t+\varepsilon_b)/2$ where $\varepsilon_{t(b)}$ is the dielectric constant of the environment above (below) the TMD monolayer, and with $r_0=\chi_{2\text{D}}/(2\kappa)$ the screening length where $\chi_{2\text{D}}$ is the 2D polarizability of the TMD layer.
\begin{figure}
\centering
\includegraphics[width=8.5cm]{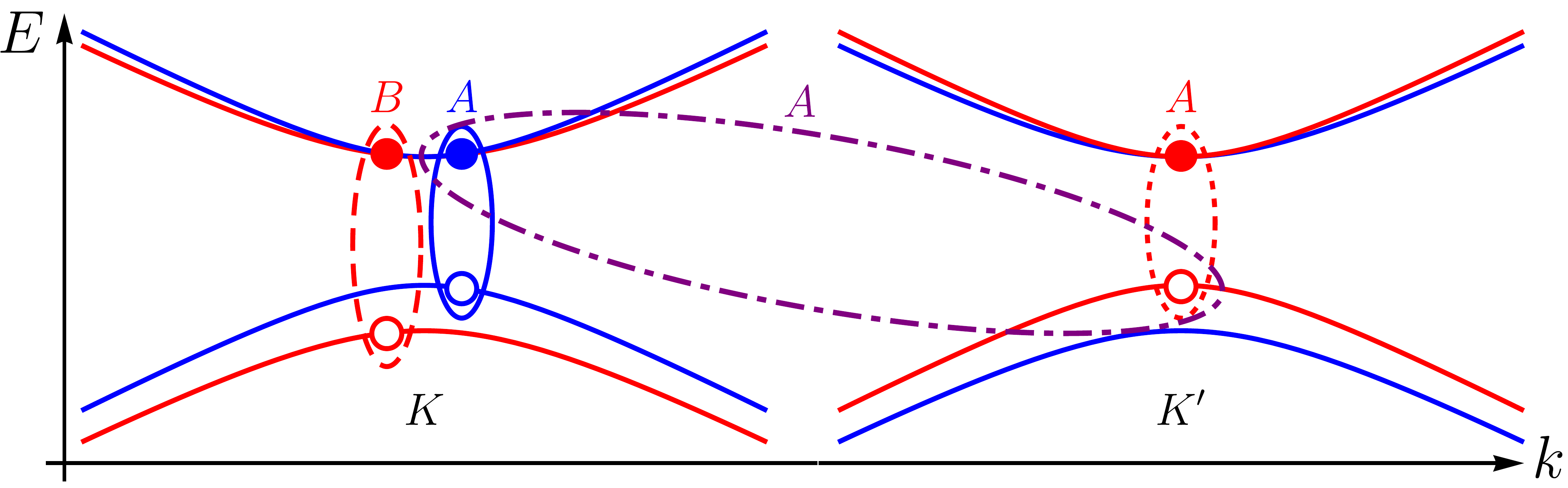}
\caption{(Color online) Schematic representation of the low-energy band structure of 2D TMDs and different kinds of excitons. Blue and red bands are spin up and spin down bands, respectively. The open and closed circles indicate holes and electrons, respectively. The blue solid ellipse and the red dotted ellipse indicate intravalley $A$ excitons in the $K$ and $K'$ valley, respectively. The red dashed ellipse indicates an intravalley $B$ exciton in the $K$ valley. The large purple dot-dashed ellipse indicates an intervalley $A$ exciton.}
\label{fig:schema}
\end{figure}
\begin{figure*}
\centering
\includegraphics[width=17cm]{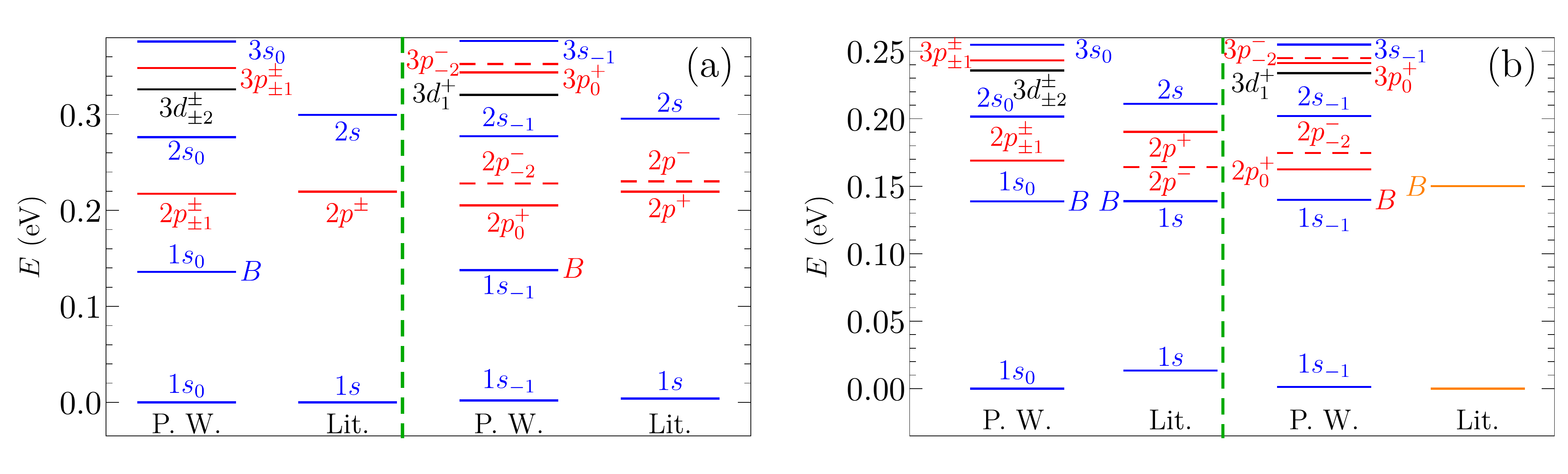}
\caption{(Color online) Intravalley ($\tau_e=-\tau_h=1$, left) and intervalley ($\tau_e=\tau_h=-1$, right) exciton energy levels for MoS$_2$ suspended in vacuum (a) and placed on a SiO$_2$ substrate ($\varepsilon_b=3.8$) (b) for $\vec{K}=\vec{0}$. The results of the present work (P. W.) are compared with results of the literature (Lit.) from Ref. [\onlinecite{levelsvacuum}] (a) and Ref. [\onlinecite{levelssio2}] (b). Our results are labeled on the figure according to $n[L]_j$, where $[L]$ represents $s$, $p$, $d$ depending on the orbital angular momentum of the dominant component of the wave function, with $s$, $p$, and $d$ excitons indicated in blue, red, and black, respectively. No labeling was given in Ref. [\onlinecite{levelssio2}] for intervalley excitons. Exciton states with negative orbital angular momentum are indicated with dashed lines. When two states with opposite orbital angular momentum are degenerate only the solid line is shown. $B$ exciton states are indicated with the letter $B$. To facilitate comparison, the energy levels are uniformly shifted downwards in energy such that the ground state (which can be either intravalley or intervalley) has zero energy.}
\label{fig:ener}
\end{figure*}

The exciton Hamiltonian is constructed in the basis $\mathcal{B}^{exc}=\{\ket{\phi^e_c}\otimes\ket{\phi^h_c},\ket{\phi^e_c}\otimes\ket{\phi^h_v},\ket{\phi^e_v}\otimes\ket{\phi^h_c},\ket{\phi^e_v}\otimes\ket{\phi^h_v}\}$ and is given by
\begin{widetext}
\begin{equation}
\label{hamtot}
H^{exc}_{\alpha}(\vec{k}^e,\vec{k}^h,r_{eh}) =
\begin{pmatrix}
-V(r_{eh}) & at(-\tau^hk_x^h-ik_y^h) & at(\tau^ek_x^e-ik_y^e) & 0 \\
at(-\tau^hk_x^h+ik_y^h) & \Delta-\lambda s^h\tau^h-V(r_{eh}) & 0 & at(\tau^ek_x^e-ik_y^e) \\
at(\tau^ek_x^e+ik_y^e) & 0 & -\Delta+\lambda s^e\tau^e-V(r_{eh}) & at(-\tau^hk_x^h-ik_y^h) \\
0 & at(\tau^ek_x^e+ik_y^e) & at(-\tau^hk_x^h+ik_y^h) & \lambda(s^e\tau^e-s^h\tau^h)-V(r_{eh})
\end{pmatrix},
\end{equation}
\end{widetext}
where the interaction term has now been added. The indices in $\alpha$ define whether the exciton is an $A$ exciton or $B$ exciton (excitons composed of a hole in the top or bottom spin-split valence band, respectively) and whether it is an \textit{intravalley} or \textit{intervalley} exciton, as illustrated in Fig. \ref{fig:schema}. The eigenvalue problem for this Hamiltonian,
\begin{equation}
\label{eigen}
H^{exc}_{\alpha}(\vec{k}^e,\vec{k}^h,r_{eh})\ket{\Psi^{exc}_{\alpha}} = E^{exc}_{\alpha}(\vec{k}^e,\vec{k}^h)\ket{\Psi^{exc}_{\alpha}},
\end{equation}
defines the exciton energy $E^{exc}_{\alpha}(\vec{k}^e,\vec{k}^h)$, from which the binding energy can be calculated through $E_{b,\alpha}^{exc}=\Delta-\lambda s^h\tau^h-E^{exc}_{\alpha}$, and the exciton eigenstate $\ket{\Psi^{exc}_{\alpha}}=\left(\ket{\phi^{e,h}_{c,c}},\ket{\phi^{e,h}_{c,v}},\ket{\phi^{e,h}_{v,c}},\ket{\phi^{e,h}_{v,v}}\right)^T$, where the subscript $\alpha$ and the superscript $exc$ have been dropped in the right hand side for notational clarity. Due to the presence of the $V(|\vec{r}_e-\vec{r}_h|)I_4$ term, the Hamiltonian does not commute with $\vec{k}^e$ nor with $\vec{k}^h$. This means that the components of the single-particle wave vectors are not good quantum numbers and should be replaced by their corresponding differential operators when solving the eigenvalue problem in the position representation. However, if we transform the single-particle coordinates to center of mass and relative coordinates,
\begin{equation}
\label{cortransf}
\vec{R} = \frac{\vec{r}_e+\vec{r}_h}{2}, \hspace{3pt} \vec{r} = \vec{r}_e-\vec{r}_h, \hspace{3pt} \vec{K} = \vec{k}^e+\vec{k}^h, \hspace{3pt} \vec{k} = \frac{\vec{k}^e-\vec{k}^h}{2},
\end{equation}
the interaction term becomes $V(r)I_4$. As a consequence, the Hamiltonian does not commute with the relative wave vector $\vec{k}$ but does commute with the center of mass momentum $\vec{K}$. Therefore, $\vec{K}$ is a conserved quantity and its components are good quantum numbers. Note that we have written the above definitions as a function of the single-particle momenta $\vec{k}$ which are relative with respect to the valley momentum $\tau\vec{D}$, i.e. $\vec{q}^i=\vec{k}^i-\tau^i\vec{D}^i$ with $\vec{q}^i$ the absolute momentum in the Brillouin zone. This means that in our coordinates the center of mass momentum of an intervalley exciton can still be zero, i.e. when both the electron and hole are located at their respective band extrema, even though the absolute center of mass momentum is $\pm2\vec{K}$ (or $\mp\vec{K}$ when reduced to the first Brillouin zone). At this point the exciton eigenvalue equation contains two variables, i.e. the two components of the relative position vector. We can try to separate these two components, in polar coordinates, by exploiting the fact that the single-electron Hamiltonian \eqref{singelhame} commutes with the angular momentum operator
\begin{equation}
\label{singleangular}
\frac{1}{\hbar}J_{z,\tau_e}^e(\vec{k}^e) = \left(x_ek_y^e-y_ek_x^e\right)I_2 + \frac{\tau_e}{2}\sigma_z = \frac{1}{i}\frac{\partial}{\partial\varphi_e}I_2 + \frac{\tau_e}{2}\sigma_z,
\end{equation}
where the first and second term correspond to the contributions from the orbital angular momentum and the pseudospin, respectively. Using $(A\otimes B)(C\otimes D)=(AC)\otimes(BD)$ and $\partial_{\varphi_e}V(r_{eh})=-\partial_{\varphi_h}V(r_{eh})$ it can be shown that the total exciton angular momentum operator
\begin{equation}
\label{excitonangular}
\begin{split}
\frac{1}{\hbar}J_{z,\tau_e,\tau_h}^{exc} &= \frac{1}{\hbar}J_{z,\tau_e}^e(\vec{k}^e)\otimes I_2 + I_2\otimes\frac{1}{\hbar}J_{z,-\tau_h}^e(\vec{k}^h) \\
&= \left(xk_y-yk_x+XK_y-YK_x\right)I_4 \\
&\hspace{10pt}+\frac{1}{2}\text{diag}\left(\tau_e-\tau_h,\tau_e+\tau_h,-\tau_e-\tau_h,-\tau_e+\tau_h\right)
\end{split}
\end{equation}
commutes with the exciton Hamiltonian \eqref{hamtot}. Note that the separate electron, hole, relative, and center of mass angular momentum operators all do not commute with the exciton Hamiltonian. In the single-band Schr\"odinger-like model, which is often used in the literature, the latter two angular momentum operators do commute with the exciton Hamiltonian, but this is prevented in the multi-band Dirac model due to the coupling between the momentum and the pseudospin.

\section{Exact solution for \texorpdfstring{$\vec{K}=\vec{0}$}{TEXT} excitons}
\label{sec:K0}

Let us first consider the simplest case in which $\vec{K}=\vec{0}$, i.e. excitons with no translational kinetic energy. The center of mass orbital angular momentum now vanishes and as a result the eigenvalues of the exciton angular momentum operator \eqref{excitonangular} are all non-degenerate and the eigenstates are eigenstates of the exciton Hamiltonian \eqref{hamtot} as well. This allows us to write the exciton wave function as
\begin{equation}
\label{ansatz}
\Psi_{\alpha}^{exc}(r) =
\begin{pmatrix}
\phi_{c,c}^{e,h}(r)e^{i\left(j-\frac{1}{2}(\tau_e-\tau_h)\right)\varphi} \\
i\phi_{c,v}^{e,h}(r)e^{i\left(j-\frac{1}{2}(\tau_e+\tau_h)\right)\varphi} \\
i\phi_{v,c}^{e,h}(r)e^{i\left(j+\frac{1}{2}(\tau_e+\tau_h)\right)\varphi} \\
\phi_{v,v}^{e,h}(r)e^{i\left(j+\frac{1}{2}(\tau_e-\tau_h)\right)\varphi}
\end{pmatrix},
\end{equation}
with $j$ the angular quantum number (which needs to be an integer in order to satisfy single valuedness of the wave function) and where the diagonal nature of the exciton angular momentum operator allowed us to include separate prefactors (in this case a factor $i$ in the second and third component) for our convenience. Using the above ansatz and transforming to relative coordinates, the exciton eigenvalue problem \eqref{eigen} in position representation becomes
\begin{widetext}
\begin{equation}
\label{excdif4}
\begin{cases}
\tau_h\left(\frac{\partial}{\partial r}-\frac{\tau_h}{r}\left(j-\frac{1}{2}\tau_{eh}^+\right)\right)\phi_{c,v}^{e,h}(r)+\tau_e\left(\frac{\partial}{\partial r}+\frac{\tau_e}{r}\left(j+\frac{1}{2}\tau_{eh}^+\right)\right)\phi_{v,c}^{e,h}(r) = \frac{1}{at}\left(E_{\alpha}^{exc}+V(r)\right)\phi_{c,c}^{e,h}(r) \\
\tau_h\left(\frac{\partial}{\partial r}+\frac{\tau_h}{r}\left(j-\frac{1}{2}\tau_{eh}^-\right)\right)\phi_{c,c}^{e,h}(r)+\tau_e\left(\frac{\partial}{\partial r}+\frac{\tau_e}{r}\left(j+\frac{1}{2}\tau_{eh}^-\right)\right)\phi_{v,v}^{e,h}(r) = -\frac{1}{at}\left(E_{\alpha}^{exc}+V(r)-\Delta+\lambda s^h\tau^h\right)\phi_{c,v}^{e,h}(r) \\
\tau_e\left(\frac{\partial}{\partial r}-\frac{\tau_e}{r}\left(j-\frac{1}{2}\tau_{eh}^-\right)\right)\phi_{c,c}^{e,h}(r)+\tau_h\left(\frac{\partial}{\partial r}-\frac{\tau_h}{r}\left(j+\frac{1}{2}\tau_{eh}^-\right)\right)\phi_{v,v}^{e,h}(r) = -\frac{1}{at}\left(E_{\alpha}^{exc}+V(r)+\Delta-\lambda s^e\tau^e\right)\phi_{v,c}^{e,h}(r) \\
\tau_e\left(\frac{\partial}{\partial r}-\frac{\tau_e}{r}\left(j-\frac{1}{2}\tau_{eh}^+\right)\right)\phi_{c,v}^{e,h}(r)+\tau_h\left(\frac{\partial}{\partial r}+\frac{\tau_h}{r}\left(j+\frac{1}{2}\tau_{eh}^+\right)\right)\phi_{v,c}^{e,h}(r) = \frac{1}{at}\left(E_{\alpha}^{exc}+V(r)-\lambda(s^e\tau^e-s^h\tau^h)\right)\phi_{v,v}^{e,h}(r)
\end{cases},
\end{equation}
\end{widetext}
which we solve numerically `exact' using the finite element method with $\tau_{eh}^{\pm}=\tau_e\pm\tau_h$. In the remainder of this work we will only consider optically created charge carriers, i.e. $s^e\tau^e=s^h\tau^h=1\ (-1)$ for $A$ ($B$) excitons.
\begin{table}
\centering
\caption{Binding energy (meV) for different $\vec{K}=\vec{0}$ intravalley exciton states in different TMDs for different substrates. We used $\varepsilon_b=3.8$ and $\varepsilon_t=1$ for SiO$_2$ with vacuum above the TMD and $\varepsilon_b=\varepsilon_t=4.4$ for encapsulating hBN.}
\begin{tabular}{c c c c c c c c c}
\hline
\hline
 & Substrate & $1s$ & $2p$ & $2s$ & $3d$ & $3p$ & $3s$ \\
\hline
\hline
Mo$\text{S}_2$ & Vacuum & 539 & 321 & 262 & 212 & 190 & 163 \\
 & SiO$_2$ & 308 & 139 & 107 & 73 & 65 & 54 \\
 & hBN & 183 & 61 & 46 & 26 & 25 & 20 \\
\hline
MoS$\text{e}_2$ & Vacuum & 472 & 291 & 241 & 199 & 179 & 154 \\
 & SiO$_2$ & 280 & 135 & 104 & 74 & 66 & 55 \\
 & hBN & 172 & 63 & 48 & 28 & 26 & 22 \\
\hline
W$\text{S}_2$ & Vacuum & 506 & 283 & 226 & 176 & 157 & 132 \\
 & SiO$_2$ & 272 & 110 & 83 & 52 & 48 & 39 \\
 & hBN  & 152 & 44 & 34 & 17 & 17 & 14 \\
\hline
WS$\text{e}_2$ & Vacuum & 458 & 265 & 214 & 170 & 152 & 129 \\
 & SiO$_2$ & 254 & 108 & 82 & 54 & 49 & 40 \\
 & hBN & 146 & 45 & 34 & 19 & 18 & 15 \\
\hline
\hline
\end{tabular}
\label{table:bindtable}
\end{table}

The results for the exciton energy levels of MoS$_2$ are shown in Fig. \ref{fig:ener}. We see that the energy difference between the energy levels is larger when the material is suspended in vacuum as compared to when it is placed on a substrate, which is due to the stronger interactions in the former case. We compare our results with those of Refs. [\onlinecite{levelsvacuum}] and [\onlinecite{levelssio2}], which use a first principles and a tight binding formulation of the Bethe-Salpeter equation approach, respectively, and indeed confirm that $E_{2p}<E_{2s}$, which was also experimentally found in Ref. [\onlinecite{spd}] for WS$_2$. Furthermore, we find that $E_{3d}<E_{3p}<E_{3s}$, which was previously theoretically predicted in Ref. [\onlinecite{spd}] for WS$_2$.

Intravalley $\vec{K}=\vec{0}$ excitons with angular momenta $\pm j$ are degenerate, which is in agreement with Ref. [\onlinecite{levelsvacuum}] but not with Ref. [\onlinecite{levelssio2}]. This degeneracy is broken for intervalley excitons, again in agreement with Ref. [\onlinecite{levelsvacuum}], but is restored when taking opposite intervalley excitons into account, i.e. $(\tau,j)$ and $(-\tau,-j)$ excitons are degenerate. This may suggest that the (non-)degeneracy between the states with opposite $j$ arises from the coupling between the exciton angular momentum and the exciton Berry curvature. The single-particle Berry curvature is opposite in the conduction and valence band, as well as in the two valleys\cite{theory1}. This means that the total Berry curvature for intravalley excitons is zero and therefore it can not couple with the exciton angular momentum. For intervalley excitons, however, the total Berry curvature is non-zero and opposite for the two opposite intervalley excitons. This causes a valley-opposing splitting between intervalley excitons with opposite $j$, which explains the degeneracy between ($\tau,j$) and ($-\tau,-j$) exciton states. The non-degeneracy between opposite $j$ intravalley excitons found in Ref. [\onlinecite{levelssio2}] could be explained by many-body Berry curvature effects\cite{berrysplit}, which are not taken into account in the present work.

Our results for MoS$_2$ show that the lowest energy intravalley exciton has slightly lower energy than the lowest energy intervalley exciton, which again agrees with Ref. [\onlinecite{levelsvacuum}] but not with Ref. [\onlinecite{levelssio2}]. However, whether the intravalley or intervalley exciton has lowest energy will also depend on the effect of exchange interactions. We do not take this effect into account and can therefore not give a definite answer on which type of exciton has the lowest ground state energy. However, the exact strength of the exchange interactions is difficult to predict and therefore even when including this effect it is still practically impossible to estimate which exciton has the lowest ground state energy\cite{levelssio2}.

Our most remarkable result is the ordering of the intervalley exciton energy levels with different $j$, with as most striking example the fact that we find that the ground state has angular momentum $j=-1$. The reason for this is related to the orbital angular momenta of the different components of the total exciton wave function. The second component, which corresponds to an exciton consisting of an electron in the conduction band and a hole in the valence band, is the most dominant one. As can be seen from Eq. \eqref{ansatz}, the dominant component of an intravalley exciton ($\tau_e=-\tau_h$) with total angular momentum $j$ also has orbital angular momentum $j$. For an intervalley exciton ($\tau_e=\tau_h=\tau$), however, the dominant component of an exciton with angular momentum $j$ has orbital angular momentum $j-\tau$. Therefore, the total wave function of an intervalley exciton with angular momentum $j=\tau$ resembles that of an $s$-like state and thus has the lowest energy. Looking at Eq. \eqref{excitonangular}, this remarkable result can be interpreted as the exciton having approximately zero orbital angular momentum (and hence an $s$-like wave function, even though this is not a good quantum number) but non-zero contribution from the pseudospin of the electron and hole, which cancels for intravalley excitons but adds up for intervalley excitons. This is why we have labeled the energy levels in Fig. \ref{fig:ener} according to the orbital angular momentum of the dominant component of the exciton wave function. In the remainder of the text the subscript $j$ in these labels will be omitted for intravalley excitons, as in this case the total angular momentum and the approximate orbital angular momentum are equal. The authors of Ref. [\onlinecite{levelsvacuum}] find that the intervalley exciton ground state is a $1s$-state, although the origin of their angular momentum labeling is not entirely clear, whereas in Ref. [\onlinecite{levelssio2}] no statement is made about the angular momentum of the intervalley exciton states.
\begin{table}
\centering
\caption{Binding energy (meV) for different $\vec{K}=\vec{0}$ intravalley exciton states in different TMDs for different substrates as found in the literature. Results for MoS$_2$ are theoretical whereas the results for WS$_2$ and WSe$_2$ are experimental.}
\begin{tabular}{c c c c c c c c c c}
\hline
\hline
 & Substrate & Ref. & $1s$ & $2p$ & $2s$ & $3d$ & $3p$ & $3s$ \\
\hline
\hline
Mo$\text{S}_2$ & Vacuum & [\onlinecite{levelsvacuum}] & 614 & 395 & 315 & - & - & - \\
 & SiO$_2$ & [\onlinecite{levelssio2}] & 301 & 150(-)/125(+) & 99 & - & - & - \\
\hline
W$\text{S}_2$ & SiO$_2$ & [\onlinecite{chernikov}] & 320 & - & 156 & - & - & 96 \\
 & SiO$_2$ & [\onlinecite{spd}] & 621 & 423 & 314 & 266 & 205 & 136 \\
\hline
WS$\text{e}_2$ & SiO$_2$ & [\onlinecite{he}] & 364 & - & 199 & - & - & 133 \\
 & SiO$_2$ & [\onlinecite{levelswse2}] & 650 & 500 & 500 & - & 370 & 370 \\
\hline
\hline
\end{tabular}
\label{table:reftable}
\end{table}

In Table \ref{table:bindtable} we give the binding energy of different exciton states in different TMDs, for which we used the parameters given in Table III of Ref. [\onlinecite{interlayer}]. The results show that, for all exciton states, the binding energy is largest in MoS$_2$ and smallest in WSe$_2$. Remarkably, the binding energy is larger in MoSe$_2$ than in WS$_2$ for all states except for the ground state. The presence of a substrate significantly decreases the exciton binding energy. Results from the literature are summarized in Table \ref{table:reftable}. In the case of MoS$_2$, for which the results from the literature are theoretical, the agreement with our results is good, differing at most 17\%. In the case of WS$_2$ and WSe$_2$ experimental results are available (except for the $2s$-, $3d$-, and $3s$-states of Ref. [\onlinecite{spd}] which are theoretical) and the agreement with our results is less satisfactory, differing at least 15\% and at most a factor 9 (with the $3s$-state of Ref. [\onlinecite{levelswse2}]). The results of Refs. [\onlinecite{spd}] and [\onlinecite{levelswse2}] in particular are remarkable. For the $1s$-state they obtain binding energies which are significantly larger than the range of commonly accepted theoretical ground state exciton binding energies in vacuum, even though in both works a SiO$_2$ substrate is used which should reduce the binding energy. However, we see that the difference in energy between the $1s$- and $2s$-state found in Ref. [\onlinecite{levelswse2}] is 13\% smaller than our result, whereas the result found in Ref. [\onlinecite{spd}] is 38\% larger than our result. This may indicate that in Ref. [\onlinecite{levelswse2}] the band gap is overestimated, which was already suggested in the manuscript itself, whereas in Ref. [\onlinecite{spd}] the sample may have been locally detached from the substrate. Another possible explanation for the discrepancy between these results and our results is substrate surface roughness, which can influence experimental measurements but is very difficult to model theoretically.
\begin{table}
\centering
\caption{Average interparticle distance (nm) for different $\vec{K}=\vec{0}$ intravalley exciton states in different TMDs for different substrates. We used $\varepsilon_b=3.8$ and $\varepsilon_t=1$ for SiO$_2$ with vacuum above the TMD and $\varepsilon_b=\varepsilon_t=4.4$ for encapsulating hBN.}
\begin{tabular}{c c c c c c c c c}
\hline
\hline
 & Substrate & $1s$ & $2p$ & $2s$ & $3d$ & $3p$ & $3s$ \\
\hline
\hline
Mo$\text{S}_2$ & Vacuum & 1.00 & 2.05 & 2.97 & 3.43 & 4.39 & 5.53 \\
 & SiO$_2$ & 1.11 & 2.57 & 3.86 & 4.87 & 6.31 & 8.09 \\
 & hBN & 1.27 & 3.47 & 5.22 & 7.49 & 9.52 & 11.99 \\
\hline
MoS$\text{e}_2$ & Vacuum & 1.04 & 2.11 & 3.05 & 3.48 & 4.46 & 5.59 \\
 & SiO$_2$ & 1.14 & 2.57 & 3.83 & 4.71 & 6.12 & 7.83 \\
 & hBN & 1.28 & 3.32 & 5.02 & 6.92 & 8.85 & 11.27 \\
\hline
W$\text{S}_2$ & Vacuum & 1.23 & 2.56 & 3.75 & 4.42 & 5.69 & 7.21 \\
 & SiO$_2$ & 1.41 & 3.43 & 5.18 & 6.87 & 8.84 & 11.32 \\
 & hBN & 1.66 & 4.93 & 7.36 & 11.33 & 14.05 & 17.52 \\
\hline
WS$\text{e}_2$ & Vacuum & 1.27 & 2.62 & 3.82 & 4.45 & 5.72 & 7.23 \\
 & SiO$_2$ & 1.43 & 3.40 & 5.12 & 6.62 & 8.56 & 10.97 \\
 & hBN & 1.66 & 4.73 & 7.11 & 10.58 & 13.24 & 16.65 \\
\hline
\hline
\end{tabular}
\label{table:disttable}
\end{table}

In Table \ref{table:disttable} we give the average interparticle distance of different exciton states in different TMDs, which are calculated from
\begin{equation}
\label{dist}
\braket{r_{eh}^{\alpha}} = 2\pi\int_0^{\infty}r^2C_{eh}^{\alpha}(r)dr,
\end{equation}
where the electron-hole correlation function is defined as
\begin{equation}
\label{corr}
C_{eh}^{\alpha}(\vec{r}) = \braket{\Psi_{\alpha}^{exc}|\delta(\vec{r}_e-\vec{r}_h-\vec{r})|\Psi_{\alpha}^{exc}}.
\end{equation}
The average interparticle distances show the opposite behavior as compared to the binding energies, as can be expected. We see that the average interparticle distance is mostly determined by the type of transition metal in the TMD, while changing the chalcogen atom between S and Se has very little influence on the interparticle distance. Furthermore, in vacuum the average interparticle distance in MoS$_2$ (WS$_2$) is slightly smaller than that in MoSe$_2$ (WSe$_2$) for all states, while in the presence of a substrate this behavior holds for the lowest states, whereas for higher excited states the opposite is true.
\\
\section{Approximate solution for \texorpdfstring{$\vec{K}\neq\vec{0}$}{TEXT} excitons}
\label{sec:Kn0}

When $\vec{K}\neq\vec{0}$, the center of mass orbital angular momentum $\vec{R}\times\hbar\vec{K}$ can be non-zero and the eigenvalues of the total exciton angular momentum operator are in general given by the sum of the relative and center of mass quantum numbers $j=j_r+j_R$, both of which are integers, meaning that these eigenvalues are all infinitely degenerate and that the eigenstates are not necessarily eigenstates of the exciton Hamiltonian. In order to find the common eigenstates of the angular momentum and the Hamiltonian the latter would have to be diagonalized in the infinite dimensional subspace spanned by all the angular momentum eigenstates corresponding to a given eigenvalue $j$. This is practically impossible and as such this prevents us from separating the angular problem from the radial one. Since the momentum-pseudospin coupling lies at the heart of this problem, this can be resolved by decoupling the exciton eigenvalue equation to a single equation following a procedure analogous to earlier works \cite{decouple1,decouple2,decouple3,decouple4}, which gives
\begin{widetext}
\begin{equation}
\label{excdifK}
\begin{split}
\bigg(&-\frac{2a^2t^2}{E^{exc}_{\alpha}+V(r)}\left(\frac{\partial^2}{\partial r^2}+\frac{1}{r}\frac{\partial}{\partial r}+\frac{1}{r^2}\frac{\partial^2}{\partial\varphi^2}-\frac{K^2}{4}\right)-V(r)+\Delta-\lambda s^h\tau^h-a^2t^2\left(\frac{\partial}{\partial r}\frac{1}{E^{exc}_{\alpha}+V(r)}\right)\bigg\{2\frac{\partial}{\partial r}+i\frac{\tau_e+\tau_h}{r}\frac{\partial}{\partial\varphi} \\
&+\delta_{\tau^h,-\tau^e}\tau^h\left[K_y\cos(\varphi)-K_x\sin(\varphi)\right]\bigg\}\bigg)\phi_{c,v}^{e,h}(r,\varphi) = E^{exc}_{\alpha}\phi_{c,v}^{e,h}(r,\varphi).
\end{split}
\end{equation}
\end{widetext}
In this approximate decoupling the kinetic energy of the particles is assumed to be small compared to the band gap and the total exciton energy, which is a good approximation for 2D TMDs. The disadvantages of using this equation are the fact that it needs to be solved self-consistently and that the other three components of the exciton wave function still need to be calculated explicitly after solving this equation, whereas a numerical solution of the coupled set of equations \eqref{excdif4} immediately yields all four components. The last two terms in this equation, which only appear for intravalley excitons and not for intervalley ones, still prevent us from separating the angular and the radial part. In principle we could treat these terms within perturbation theory. In such a case the angular part of the zeroth order wave function is simply given by $\text{exp}(ij\varphi)$ with $j$ an integer quantum number. This implies that these terms give no contribution in first order perturbation theory, whereas in second order perturbation theory they only couple states whose angular momentum quantum numbers differ by $\pm1$. We can therefore assume that the total contribution of these two terms will be negligibly small and we will neglect them in the remainder of our calculations and thus assume the angular part of the wave function to be given by $\text{exp}(ij\varphi)$.

\begin{figure}
\centering
\includegraphics[width=8.5cm]{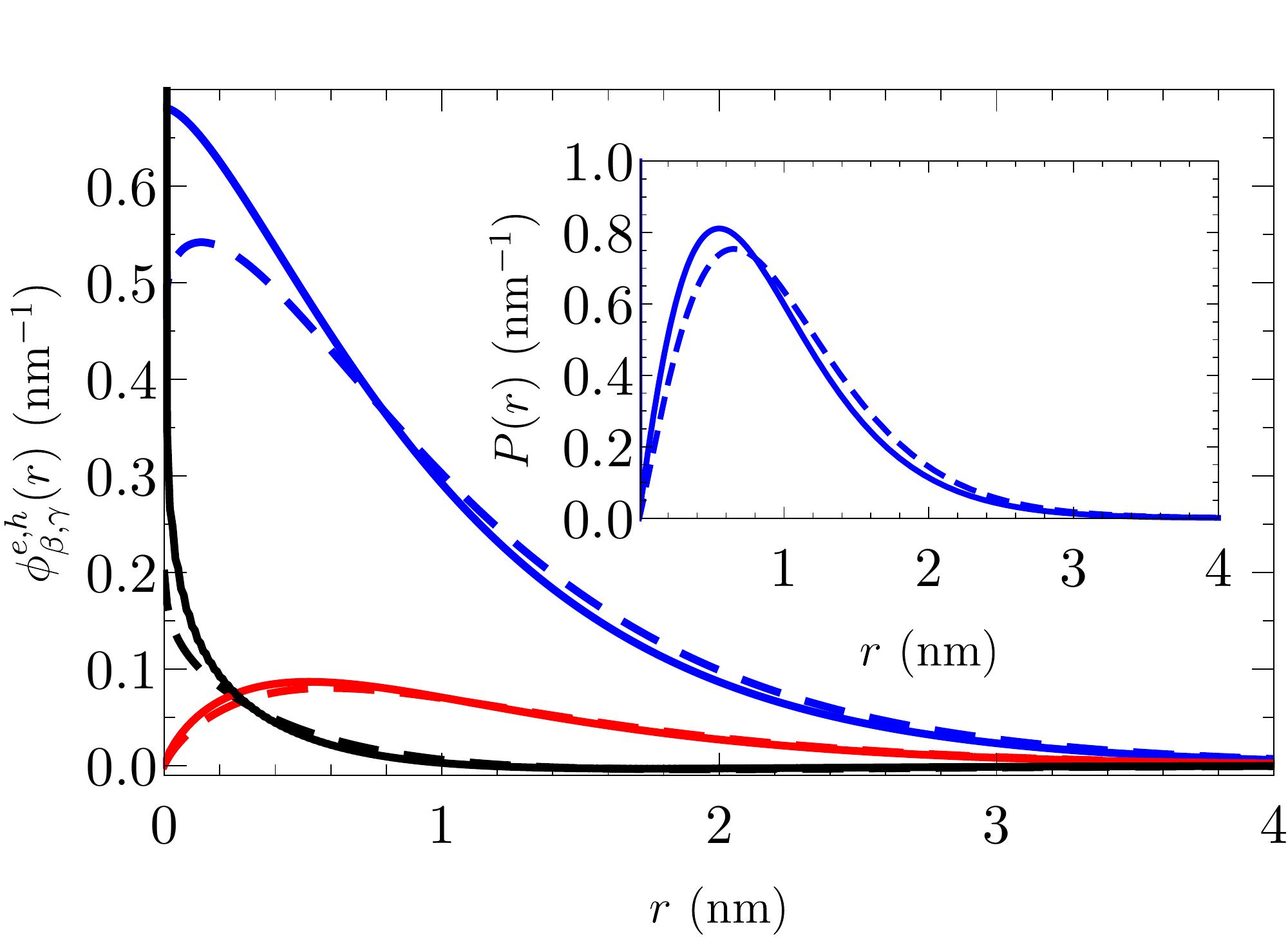}
\caption{(Color online) Different components ($\beta=c,\gamma=v$: blue curve, $\beta=c$, $\gamma=c$ and $\beta=v$, $\gamma=v$: red curve, $\beta=v$, $\gamma=c$: black curve) of the $\vec{K}=\vec{0}$ intravalley $1s$-state wave function for excitons in MoS$_2$ suspended in vacuum obtained from Eq. \eqref{excdifK} (solid) and Eq. \eqref{excdif4} (dashed). The blue and black curves are $s$-like and have zero pseudospin, the red curves are $p$-like and have pseudospin $\pm1$. The total angular momentum of this state is $j=0$. The inset shows the corresponding total radial probability distribution for the two cases.}
\label{fig:golf}
\end{figure}

Using the above equation for a $\vec{K}=\vec{0}$ intravalley exciton in MoS$_2$ suspended in vacuum, we find a binding energy of 556 meV, 274 meV, and 330 meV for the $1s$-, $2s$-, and $2p$-state, respectively. As a comparison, using equation \eqref{excdif4} we find 539 meV, 262 meV, and 321 meV, respectively, which amounts to a difference of 3-4\% between the results of the two equations. The four components of the $1s$-state exciton wave function obtained by the two different methods are shown in Fig. \ref{fig:golf}. It is clear that the second component of the exciton wave function (blue curves), which represents the contribution of an exciton composed of an electron in the conduction band and a hole in the valence band, is significantly larger than the other three components. This component as well as the third component (black curves) show $s$-like behavior, whereas the first and fourth component (red curves) show $p$-like behavior. This is in agreement with Eq. \eqref{ansatz}, in which for $j=0$ and $\tau^e=-\tau^h=1$ the second and third component have zero orbital angular momentum and the first (fourth) component has orbital angular momentum $-1$ $(1)$. The largest difference between the two methods is found in the second component (the blue curves in Fig. \ref{fig:golf}) for small $r$. The wave function obtained from Eq. \eqref{excdif4} has a maximum at small non-zero $r$ while the solution obtained from Eq. \eqref{excdifK} has its maximum in the origin. This additional curvature of the wave function for the former leads to a higher energy and as such a lower binding energy, in agreement with the values mentioned above. The total radial probability distributions as obtained from the two equations are in good agreement, with the probability distribution obtained from Eq. \eqref{excdif4} being slightly more spread out, which is in agreement with the lower binding energy which was found using this equation.

The exciton energy shows parabolic dependence as a function of the center of mass momentum at low energy. From the curvature we obtain the total exciton mass through the expression
\begin{equation}
\label{excmass}
M = \frac{\hbar^2}{m_0}\left(\frac{\partial^2E^{exc}_{\alpha}(K)}{\partial K^2}\Bigr|_{K=0}\right)^{-1},
\end{equation}
with $m_0$ the free electron mass. We find $M=1.14m_0$, $M=1.24m_0$, $M=0.70m_0$, and $M=0.77m_0$ for $1s$-state excitons in, respectively, MoS$_2$, MoSe$_2$, WS$_2$, and WSe$_2$ suspended in vacuum. This is in good agreement with the values derived from the literature\cite{excmass} for equal electron and hole masses, i.e. $M=2m_{e/h}$, for which we find, respectively, $M=1.00m_0$, $M=1.08m_0$, $M=0.64m_0$, and $M=0.68m_0$, which is 9-13\% smaller than our calculated exciton masses. The effective mass of the single-particle energy spectrum of Eq. \eqref{singelhame} is given by $m=\hbar^2(\Delta-\lambda)/(2a^2t^2)$, from which we find exciton masses of, respectively, $M=0.97m_0$, $M=1.09m_0$, $M=0.62m_0$, and $M=0.68m_0$, which is 11-15\% smaller than our calculated exciton masses. Furthermore, we find that these values differ very little for higher excited exciton states (which have slightly smaller masses, e.g. $M=1.12m_0$ and $M=1.09m_0$ for the $2p$- and $2s$-state in MoS$_2$, respectively) and that there is also a very weak dependence on the substrate dielectric constant (decreasing mass as a function of the dielectric constant, e.g. $M=1.09m_0$ for the $1s$-state exciton in MoS$_2$ on a SiO$_2$ substrate).

The exchange interactions couple the intravalley exciton bands originating from direct transitions in the $K$ and $K'$ valley and as such lead to a splitting of these originally degenerate bands into a parabolic lower band a linear upper band\cite{levelsvacuum,levelssio2}. This leads to a correction on the total exciton mass of the lower parabolic band, which is the ground state. Using the effective model and the parameters given in Eq. (12) of Ref. [\onlinecite{levelsvacuum}] we find a correction factor of 1.26 for the total exciton mass for MoS$_2$ suspended in vacuum. The presence of a substrate will reduce this correction factor. Similar results are expected for other TMDs.

\section{Exciton Landau levels}
\label{sec:excitonll}

\begin{figure}
\centering
\includegraphics[width=8.5cm]{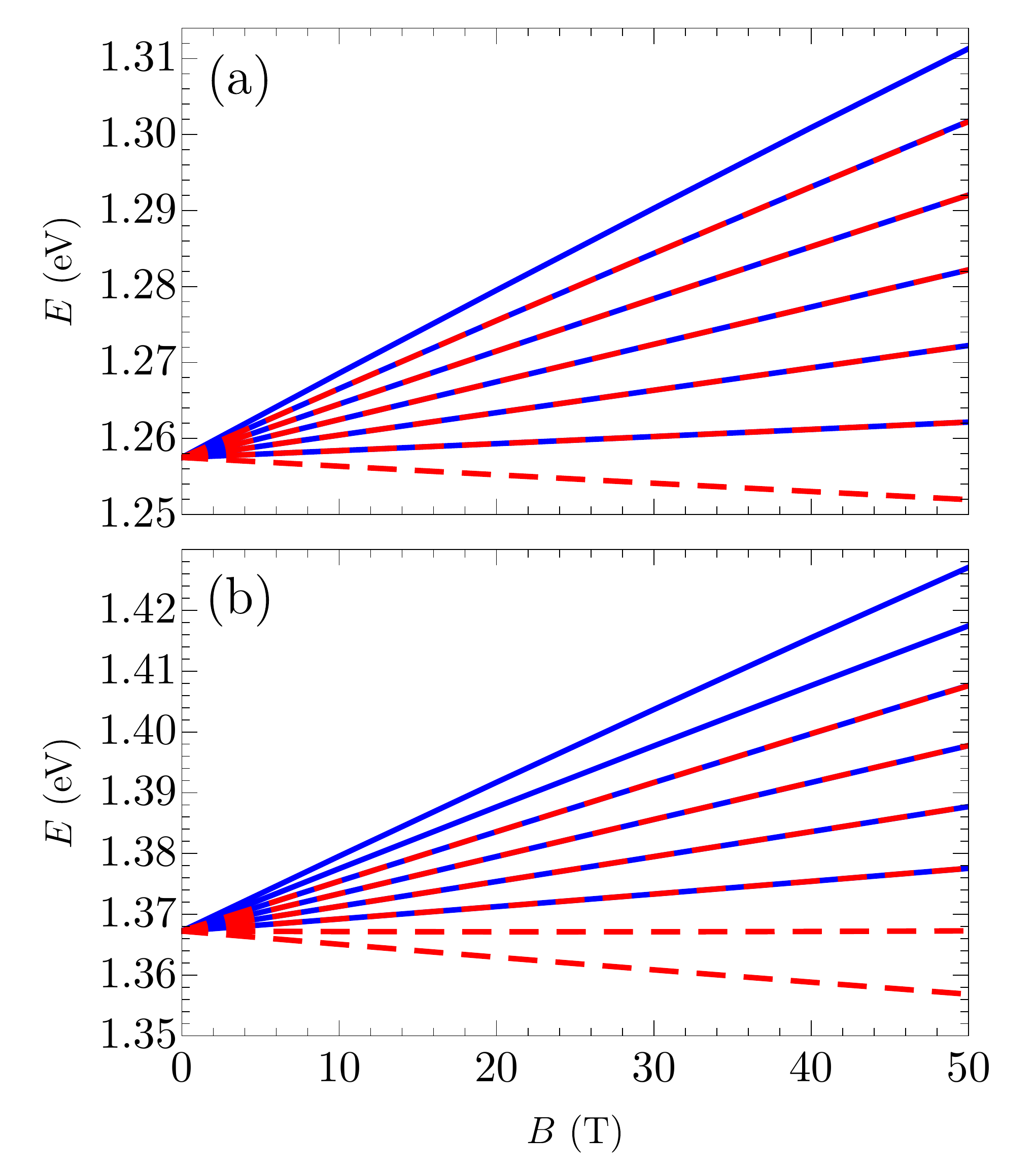}
\caption{(Color online) Six lowest Landau levels of the intravalley ($\tau_e=-\tau_h=1$) $2p$-state (a) and $3d$-state (b) for excitons in MoS$_2$ suspended in vacuum. Solid blue (red dashed) lines indicate positive (negative) relative angular momentum states.}
\label{fig:landau}
\end{figure}

\begin{figure}
\centering
\includegraphics[width=8.5cm]{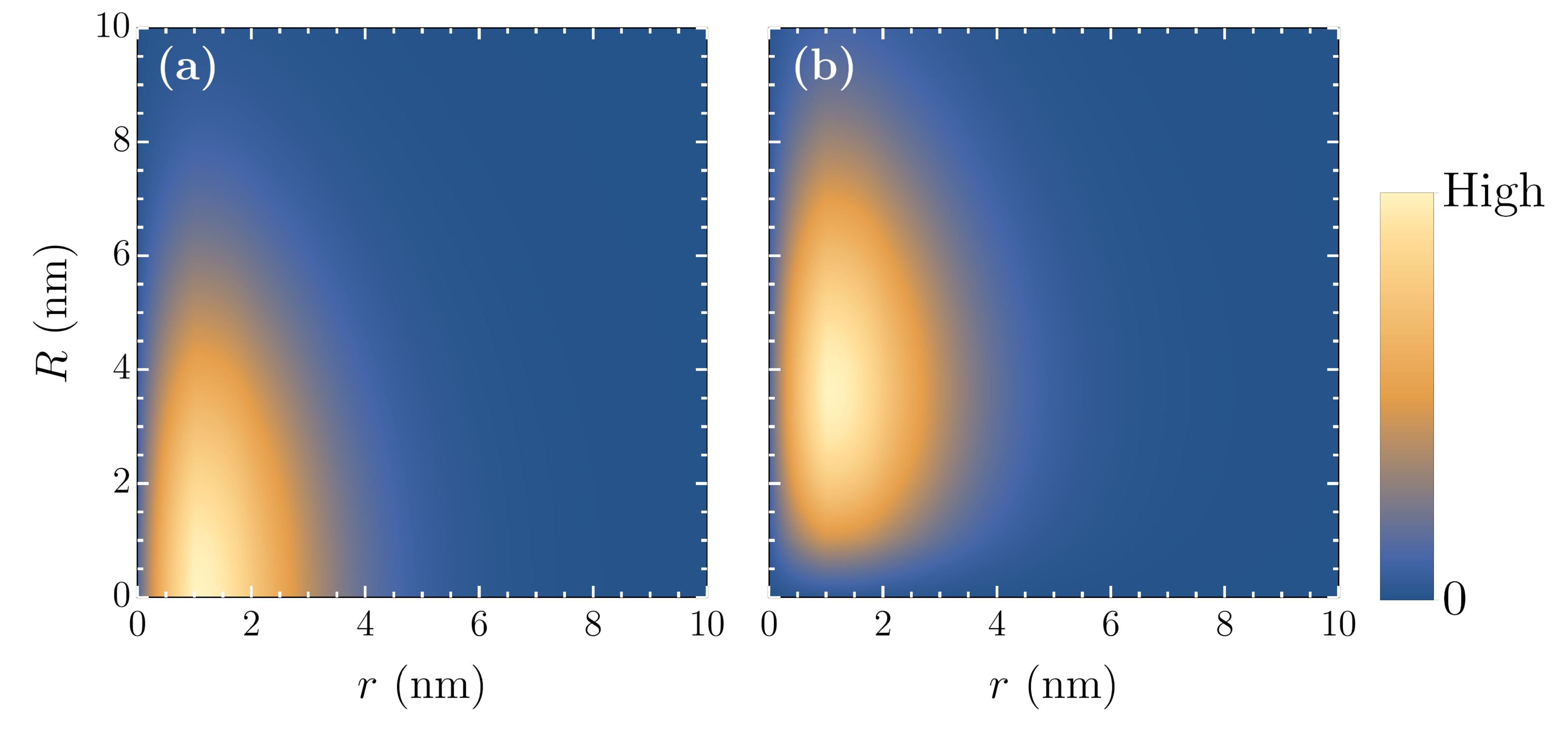}
\caption{(Color online) Dominant component of the intravalley ($\tau_e=-\tau_h=1$) $2p$-state wave function for the degenerate states with $(j_r,j_R)=(1,0)$ (a) and $(j_r,j_R)=(-1,1)$ (b) for excitons in MoS$_2$ suspended in vacuum in the presence of a magnetic field of $B=50$ T.}
\label{fig:degenplot}
\end{figure}

In the presence of a perpendicular magnetic field the wave vectors are replaced by $\vec{\Pi}=\vec{k}-q\vec{A}/\hbar$ where $\vec{A}$ is the vector potential giving rise to the magnetic field $\vec{B}=\vec{\nabla}\times\vec{A}$. Here, we choose to work in the symmetric gauge $\vec{A}=(-By/2,Bx/2,0)^T$. As a result, the components of the center of mass momentum are no longer good quantum numbers and we will again need to decouple the exciton eigenvalue equation in order to separate the angular part from the radial part, even though the total exciton angular momentum operator \eqref{excitonangular} still commutes with the (magnetic) Hamiltonian. Furthermore, we also need to take into account the spin Zeeman effect and the Zeeman effect due to the orbital angular momentum $m$ of the single-particle states around their atomic sites, i.e. $m=0$ and $m=2\tau$ for conduction and valence band states, respectively. Eventually we find
\begin{widetext}
\begin{equation}
\label{excdifB}
\begin{split}
\bigg\{&-\frac{2a^2t^2}{g(r,E^{exc}_{\alpha})}\bigg[\frac{\partial^2}{\partial r^2}+\frac{1}{r}\frac{\partial}{\partial r}+\frac{1}{r^2}\frac{\partial^2}{\partial\varphi_r^2}+\frac{1}{4}\frac{\partial^2}{\partial R^2}+\frac{1}{4}\frac{1}{R}\frac{\partial}{\partial R}+\frac{1}{4}\frac{1}{R^2}\frac{\partial^2}{\partial\varphi_R^2}-\frac{1}{16l_B^4}\left(r^2+4R^2\right)+\frac{i}{2l_B^2}\left(\frac{\partial}{\partial\varphi_r}+\frac{\partial}{\partial\varphi_R}\right) \\
&-\frac{\tau^e+\tau^h}{2l_B^2}\bigg]-a^2t^2\left(\frac{\partial}{\partial r}\frac{1}{g(r,E^{exc}_{\alpha})}\right)\bigg(2\frac{\partial}{\partial r}+i\frac{\tau_e+\tau_h}{r}\frac{\partial}{\partial\varphi_r}-(\tau^e+\tau^h)\frac{r}{4l_B^2}+\delta_{\tau^h,-\tau^e}\tau^h\bigg[\cos(\varphi_R-\varphi_r)\frac{R}{l_B^2} \\
&-i\sin(\varphi_R-\varphi_r)\frac{\partial}{\partial R}-\cos(\varphi_R-\varphi_r)\frac{i}{R}\frac{\partial}{\partial\varphi_R}\bigg]\bigg)-V(r)+\Delta-\lambda s^h\tau^h\bigg\}\phi_{c,v}^{e,h}(r,R,\varphi_r,\varphi_R) = E^{exc}_{\alpha}\phi_{c,v}^{e,h}(r,R,\varphi_r,\varphi_R),
\end{split}
\end{equation}
\end{widetext}
with $l_B=\sqrt{\hbar/(eB)}$ the magnetic length and $g(r,E^{exc}_{\alpha})=E^{exc}_{\alpha}+V(r)-(s^e+s^h)\mu_BB-2\tau^h\mu_BB$ with $\mu_B$ the Bohr magneton. We used the same assumptions and approximations to arrive at this equation as we did to obtain Eq. \eqref{excdifK}. There are now three terms which only appear for intravalley excitons and not for intervalley ones and which prevent the equation from being separable into an angular and a radial part. These terms are small because they are related to magnetic field effects and we can again argue that they can be neglected because they will only contribute in second order perturbation theory, where they now only couple states whose relative (center of mass) angular momentum quantum numbers differ by $\pm1$ ($\mp1$). We will therefore assume that the angular part of the wave function is given by $\text{exp}(ij_r\varphi_r+ij_R\varphi_R)$. Furthermore, we can now also distinguish the terms (i.e. the last two on the first line of the equation) corresponding to the magnetic angular momentum and the Zeeman effect related to the intrinsic magnetic moment of the individual Bloch particles \cite{bloch1,bloch2}. The Landau levels for the exciton $2p$- and $3d$-states of MoS$_2$ are shown in Fig. \ref{fig:landau}. The Landau levels show a linear behavior as a function of the magnetic field strength and we find that they correspond qualitatively to the Landau levels of a 2D charged Schr\"odinger particle, i.e.\cite{2deg}
\begin{equation}
\label{2Dgas}
E \simeq \left(n_R+\frac{j_R+|j_R|}{2}+\frac{1}{2}\right)\hbar\omega_c,
\end{equation}
with $n_R$ the (positive integer) principal center of mass quantum number, $j_R$ the center of mass angular momentum, and with $\omega_c=2eB/M$ the center of mass cyclotron frequency, meaning that for each $n_R$ the corresponding Landau levels are infinitely degenerate for all $j_R\leq0$ and that the states defined by $(n_R,j_R)$ and $(n_R+1,j_R-1)$ are degenerate for $j_R>0$. Note that some of the lowest Landau levels decrease as a function of the magnetic field, which is a consequence of the Zeeman effect due to the orbital angular momentum of the single-particle states around their atomic sites. Furthermore we find that the magnetic field breaks the degeneracy between states with opposite relative angular momentum $\pm j_r$, which is to be expected. However, there is still a degree of degeneracy in the relative angular momentum quantum number in the sense that Landau level number $k$ of the state with relative angular momentum $j_r$ is degenerate with Landau level number $k+j_r$ of the state with opposite relative angular momentum $-j_r$. As a result, only the lowest $|j_r|$ Landau levels of the state with negative relative angular momentum are non-degenerate with the Landau levels of the state with opposite relative angular momentum. This is a remarkable result since it is not immediately clear from Eq. \eqref{excdifB} that this should be the case. The exciton wave functions of two degenerate states of which both the relative and the center of mass angular momentum quantum numbers are different are shown in Fig. \ref{fig:degenplot}. The wave function in Fig. \ref{fig:degenplot}(a) shows $s$-like behavior as a function of the center of mass coordinate whereas the wave function in Fig. \ref{fig:degenplot}(b) shows $p$-like behavior. Both wave functions show $p$-like behavior as a function of the relative coordinate. This also shows that, even for a high magnetic field strength of 50 T, the exciton wave functions are more localized as function of the relative coordinate as compared to the center of mass coordinate. Note that the exchange interactions for $p$- and $d$-states are expected to be negligible since these are proportional with the value of the exciton wave function squared in the relative coordinate origin\cite{levelssio2}. Therefore, it is to be expected that the inclusion of exchange interaction effects would have little to no effect on the results presented in this Section. For $s$-states the slopes of the Landau levels are altered by the inverse of the correction factor discussed at the end of the previous Section.

In principle it should be possible to measure these exciton Landau levels experimentally by means of photoluminescence experiments. However, due to the small energy separation between the different Landau levels, high magnetic field strengths, high laser powers, and low temperatures would be needed to try and resolve the different states. Landau level-like features were found in Ref. [\onlinecite{valley6}], although the origin of these features was not discussed.

\section{Summary and conclusion}
\label{sec:Summary and conclusion}

Different excited intra- and intervalley exciton states in different monolayer TMDs were investigated. We started from the single-particle Dirac Hamiltonian to construct a four-band exciton Hamiltonian and we solved the corresponding eigenvalue equation using the finite element method. We constructed the total exciton angular momentum operator and showed how its eigenstates can be exploited to decouple the angular part from the radial part in the exciton Hamiltonian eigenvalue equation. We calculated the exciton energy levels and found that intravalley exciton states with larger angular momentum have lower energy, i.e. $E_{2p}<E_{2s}$, $E_{3d}<E_{3p}<E_{3s}$, \ldots, which agrees with earlier theoretical and experimental findings. For intervalley excitons in the $\mp K$ valley we found that the ground state has angular momentum $j=\pm1$, which is due to the contribution from the pseudospin of the electron and hole, which cancels for intravalley excitons but not for intervalley excitons. We also calculated the exciton binding energy and average interparticle distance for different combinations of excited states, TMDs, and substrates.

Furthermore, we explained why this method of separation of variables fails in the case of finite exciton center of mass momentum or in the presence of a perpendicular magnetic field. However, we showed that it is still possible to approximately separate the variables in these cases and demonstrated good agreement with the non-approximate method in the limit of zero center of mass momentum. By calculating the exciton energy as a function of the center of mass momentum we obtained the exciton mass.

Finally, we calculated the exciton Landau levels and found that they correspond qualitatively to those of a 2D charged Schr\"odinger particle. Furthermore, the perpendicular magnetic field breaks the degeneracy between states with opposite relative angular momentum but this degeneracy is partly restored when taking into account states with higher center of mass angular momentum.

\section{Acknowledgments}

This work was supported by the Research Foundation of Flanders (FWO-Vl) through an aspirant research grant for MVDD and by the FLAG-ERA project TRANS-2D-TMD.

\end{document}